\documentclass[12pt,preprint]{aastex}

\shorttitle{LP 714-37: A triple system}
\shortauthors{Phan-Bao et al.}

\begin{document}

\title{LP 714-37: A wide pair of ultracool dwarfs actually is a 
       triple$^{\star}$}

\author{Ngoc Phan-Bao,\altaffilmark{1} Thierry Forveille, \altaffilmark{2,3}
Eduardo L. Mart\'{\i}n, \altaffilmark{4,1} Xavier Delfosse \altaffilmark{3}}

\altaffiltext{$^{\star}$}{Based on observations made at the
 Canada-France-Hawaii Telescope, 
operated by the National Research Council of Canada, the Centre National 
de la Recherche Scientifique de France and the University of Hawaii.}
\altaffiltext{1}{University of Central Florida, Dept. of Physics, 
PO Box 162385, Orlando, FL 32816-2385, USA; pngoc@physics.ucf.edu}
\altaffiltext{2}{Canada-France-Hawaii Telescope Corporation, 65-1238 
Mamalahoa Highway, Kamuela, HI 96743 USA; forveille@cfht.hawaii.edu}
\altaffiltext{3}{Laboratoire d'Astrophysique de Grenoble, Universit\'e 
J. Fourier, BP 53, 38041 Grenoble, France; delfosse@obs.ujf-grenoble.fr}
\altaffiltext{4}{Instituto de Astrof\'{\i}sica de Canarias, 
C/ V\'{\i}a L\'actea s/n, E-38200 La Laguna (Tenerife), Spain; ege@iac.es}

\begin{abstract}

LP~714-37 was identified by Phan-Bao et al. (2005) as one of the very
few wide pairs of very low mass (VLM) stars known to date, with a separation
of 33~AU. Here we present adaptive optics imaging which resolves the 
secondary of the wide pair into a tighter binary, with a projected angular
separation of 0.36\arcsec, or 7~AU. The estimated spectral types of 
LP~714-37B and LP~714-37C are M8.0 and M8.5. We discuss the implications 
of this finding for brown dwarf formation scenarios.

\end{abstract}


\section{Introduction}

Binary systems offer the only practical opportunity to measure accurate 
stellar masses, and represent a powerful test of evolutionary models and 
star formation theories. These last aspects are particularly important 
for stars at the bottom of the main sequence and brown dwarfs (BDs), whose 
physical parameters and formation mechanism are not well known. Significant
attention has recently been concentrated on binaries among ultracool dwarfs
(spectral types later than M6) both in the field (\citealt{martin99, reid01};
\citealt{close02}, 2003; \citealt{bouy03,burgasser03}; 
\citealt{forveille04}, 2005; \citealt{siegler, liu, law}) and in nearby 
young open clusters and associations (\citealt{martin98}, 2000, 2003; 
\citealt{luhman04a}; \citealt{kraus}; \citealt{stassun}; \citealt{bouy06a}).
These surveys demonstrate that wide binary systems (semi-major axis $>$ 
15~AU) are very rare amongst ultracool dwarfs, with a frequency below 1\%, 
while tighter binaries (semi-major axis $<$ 15~AU) occurs in $>$15\% of 
them. This well defined cutoff is not caused by disruptions during galactic 
dynamical encounters, which only become effective at significantly wider 
separations \citep{weinberg87}. One possible explanation lies in the
ejection model \citep{reipurth, bate02} for brown dwarf formation,
which describes them as failed stellar embryos, ejected from their parent
gas core before they have time to accrete a larger mass. Wide binaries
are not expected to survive such an ejection, and the numerical simulations
of this process suggest that it only produces close binaries (separation 
$\leq$ 10~AU; \citealt{bate02}), in rough agreement with the observations.
More recently however, a growing number of wide binaries (separation $>$ 
30~AU) has been detected \citep{luhman04a, phan-bao05, chauvin, 
luhman05, billeres, bouy06b}, suggesting that at least some VLM stars 
and BDs form through
another process. Some of these apparent binaries, however, could 
be unresolved higher multiplicity systems, with a correspondingly higher
total mass and binding energy. Ascertaining how many of them there are is
clearly important.

In this Letter, we present adaptive optics observations of one such
system, LP~714-37 \citep{phan-bao05}, prompted in part by its apparently
overluminous secondary. The new images do resolve that secondary
into a tighter pair, and demonstrate that the system contains three 
VLM stars. Sec.~2 presents the observations and the data reduction, while
Sec.~3. discusses the results in the context of the binary properties of 
wide VLM stars and BDs in general, and of the formation of LP~714-37
in particular. 

\section{Observations and data reduction}

We observed LP~714-37 at the 3.6-meter Canada-France-Hawaii
Telescope (CFHT) on 2005 October 13$^{th}$, using the CFHT Adaptive 
Optics Bonnette (AOB) and the KIR infrared camera. The AOB, also called 
PUEO after the sharp-visioned Hawaiian owl, is a general-purpose adaptive 
optics (AO) system based on F. Roddier's curvature concept 
\citep{roddier91}. It is mounted at the telescope F/8 Cassegrain focus, 
and cameras or other instruments are then attached to it 
\citep{arsenault94,rigaut94}. The atmospheric 
turbulence is analysed by a 19-element wavefront curvature sensor and 
the correction applied by a 19-electrode bimorph mirror. Modal control 
and continuous mode gain optimization \citep{gendron94,rigaut94} 
maximize the quality of the AO correction for 
the current atmospheric turbulence and guide star magnitude, and produced
well corrected images for the faint LP~714-37 system ($V=16.5$, $I=13.0$). 
For our observations a dichroic mirror diverted the visible light to the 
wavefront sensor while the KIR science camera \citep{doyon98} recorded 
infrared photons. The KIR plate scale is 0\farcs035 per pixel, for a 
total field size of $36\arcsec \times 36\arcsec$.

\begin{figure}
\vskip 1in
\hskip -0.25in
\centerline{\includegraphics[width=4in,angle=0]{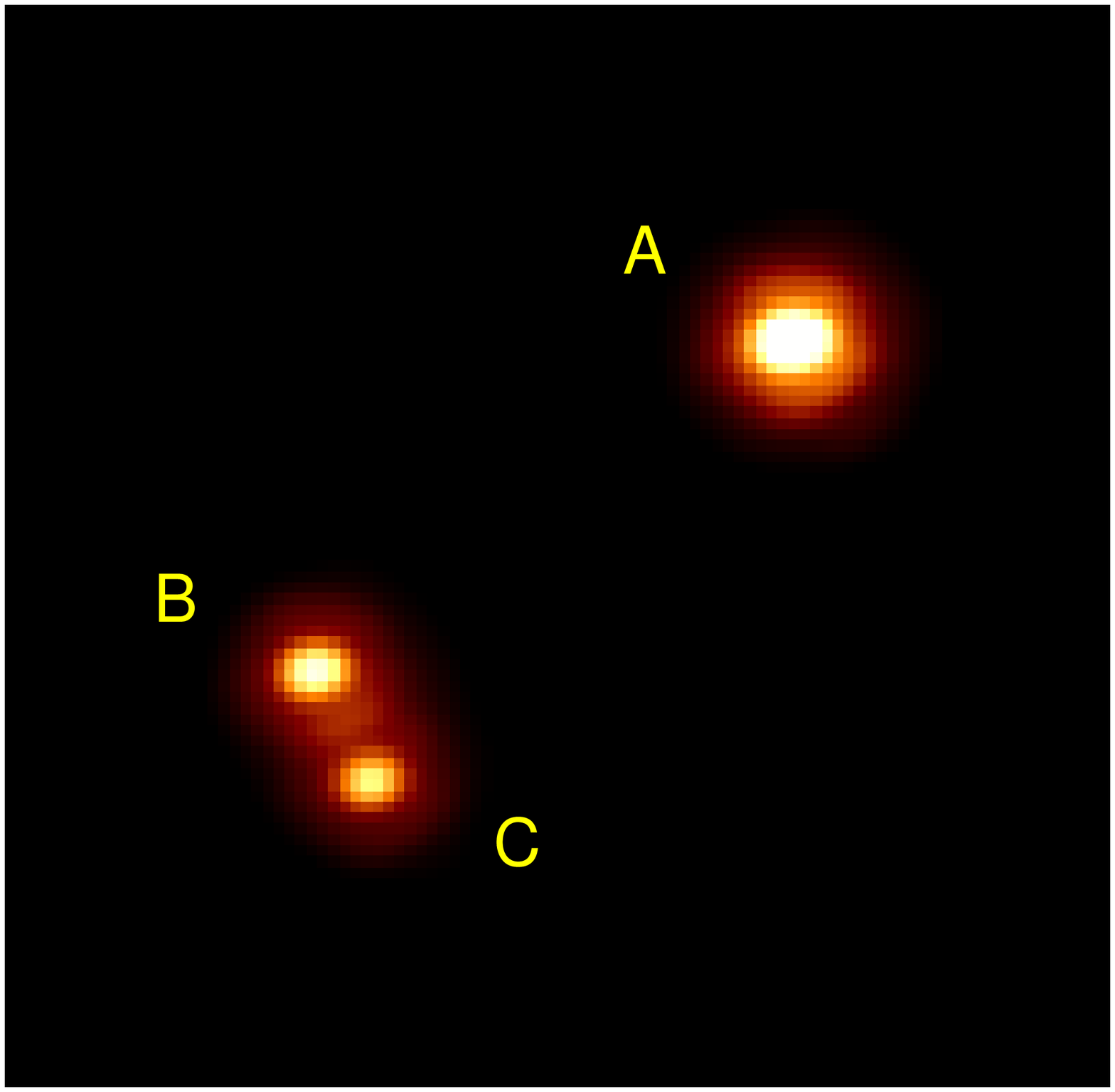}}
\caption{\normalsize 
An adaptive optics image of LP 714-37 with CFHT in the K$^{\prime}$ filter. The size of
the image is 3.5\arcsec $\times$ 3.5\arcsec, and North is up and East to the left.
\label{AOimage}}
\end{figure}

We observed LP~714-37 through a K$^{\prime}$ filter, and obtained series
of five 30s exposures at each of five dither positions. The raw images 
were median combined to produce sky frames, which were then subtracted 
from the raw data. Subsequent reduction steps included flat-fielding, 
flagging of the bad pixels, and finally shift-and-add combinations of the
corrected frames into one final image (Fig.~\ref{AOimage}) with a 12.5~min total exposure time.
Analysis of this image with SExtractor \citep{bertin} produced the relative
astrometry and relative photometry summarized in Table~\ref{tab_oa}.

\section{Discussion}

\subsection{Physical parameters of the LP~714-37 triple system}

\begin{figure}
\vskip 1in
\hskip -0.25in
\centerline{\includegraphics[width=4in,angle=0]{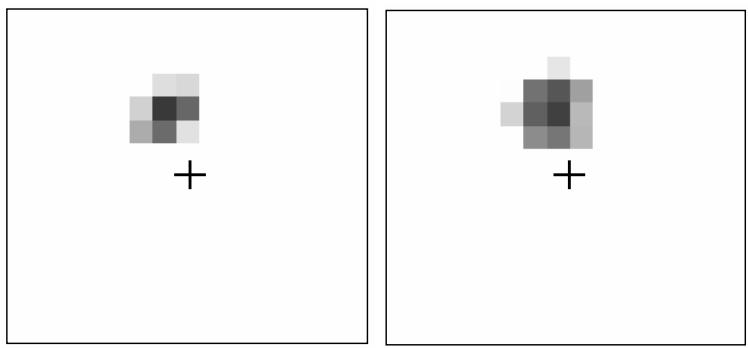}}
\caption{\normalsize 
Archival images of the LP 714-37 system: 2MASS-K (left, epoch: 1998.690 from ALADIN) and DENIS-K 
(right, epoch: 2000.896). The cross indicates the position of component C at the 
2005.784 epoch of the CFHT image. Component C would clearly be separated from LP~714-37AB
in the 2MASS-K and DENIS-K images if it were not physically associated to the system.
The size of each image is 15\arcsec $\times$ 15\arcsec, and North is up and East to the left.
\label{fig_pm}}
\end{figure}

The proper motion of the system is $\mu_{\rm \alpha}$~=~$-$117~mas/yr 
and $\mu_{\rm \delta}$~=~$-$382~mas/yr \citep{phan-bao03}, and it has 
moved by 1.95\arcsec~between the epoch of the DENIS image (2000.0) and 
our CFHT observation. Figure~\ref{fig_pm} shows no background star at 
the position of the system in the DENIS and 2MASS K images, demonstrating 
that the system is a physical triple.

\citet{phan-bao05} spectroscopically classified component~A as M5.5, with
an estimated absolute magnitude of $M_{K} = 9.11$. The relative photometry
listed in Table~\ref{tab_oa} therefore provides estimates of $M_{K}~=~10.05$ 
for LP~714-37B and $M_{K}~=~10.35$ for LP~714-37C. We note that the 
difference between the DENIS-K, 2MASS-K and K$^{\prime}$ photometry is 
very small \citep{carpenter}, and completely negligible for the purpose
of the present paper. Adding the flux of the three components, the absolute
K-band magnitude of the system is $M_{\rm K}$~=~8.5, which combines with the 
DENIS magnitude ($K_{\rm s}=9.89$) to give a photometric distance of 
18.9$\pm$2.6~pc. 
At this updated distance
the projected separations between B and C, and A and the BC barycenter,
are respectively of 6.8 and 36.1~AU.

\begin{figure}
\vskip 1in
\hskip -0.25in
\centerline{\includegraphics[width=4in,angle=-90]{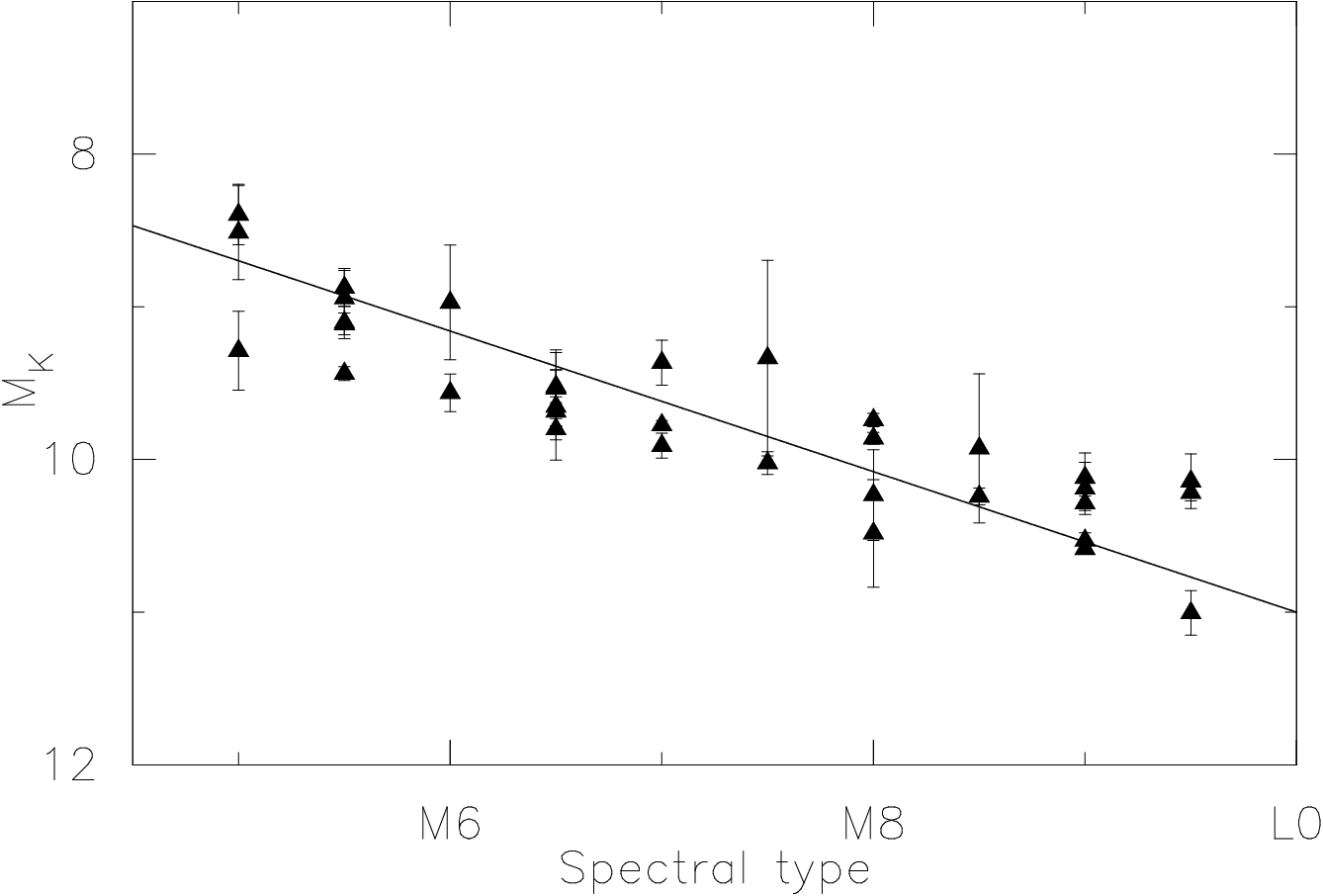}}
\caption{\normalsize 
K-band absolute magnitude as a function of spectral type; most photometric
data from \citet{phan-bao03}, \citet{dahn02}; trigonometric parallaxes from \citet{van95}, \citet{monet92},
\citet{dahn02} and references therein.
\label{MkvsSpT}}
\end{figure}

We estimate approximate spectral types for components B and C from their
absolute K-band magnitude (which itself scales back to the spectral type 
of A). A linear least-square fit to the absolute magnitudes and spectral 
types relation of 35~single M5.0-M9.5 dwarfs (Fig.~\ref{MkvsSpT}), gives
the following relation:
\begin{center}
SpT  =  2.17 $M_{\rm K}$ - 13.9, $\sigma = 0.68 $
\end{center}
where SpT is the spectral subtype, 5.0 for spectral type M5.0 and 9.5 
for spectral type M9.5. Applying this relation results in estimated 
spectral types of M8.0$\pm$0.5 and M8.5$\pm$0.5 for component B and C, 
respectively. 

The 5~Gyr K-band mass-luminosity relation of \citet{baraffe98} results
in a mass 0.08$\pm$0.01~M$_{\odot}$ for component C. All three components 
have estimated masses close to the hydrogen-burning limit \citep{chabrier},
and the total mass of LP~714-37 is $\sim$0.28~M$_{\odot}$. Table~\ref{parameters}
presents a summary of the derived physical parameters of the system.

\subsection{Could LP 714-37 have formed through the ejection process?}

Ejection models \citep{reipurth, bate02} suggest that (most) brown dwarfs 
form through the premature removal of pre-stellar cores from their parental 
molecular clouds by dynamical interactions. These models qualitatively
predict that the binary brown dwarf systems that do exist must be close 
(separation $\leq$ 10~AU), since the small binding energy of wide BD 
binaries leaves them vulnerable to disruption. More recent simulations
\citep{bate05} do produce some wide BD binary systems, when two unrelated 
objects are simultaneously ejected in the same direction. This mechanism 
however needs high density environments to work. It could thus not possibly
form the wide binaries known in TW~Hya, Cha~I, and Upper~Sco 
\citep{chauvin, luhman04a, luhman05}. A caveat, however, is that some 
apparent binaries might be unresolved triple or higher order multiple 
systems, whose additional components could boost the binding energy of the 
systems enough to allow them to survive ejection.

\begin{figure}
\vskip 1in
\hskip -0.25in
\centerline{\includegraphics[width=4in,angle=-90]{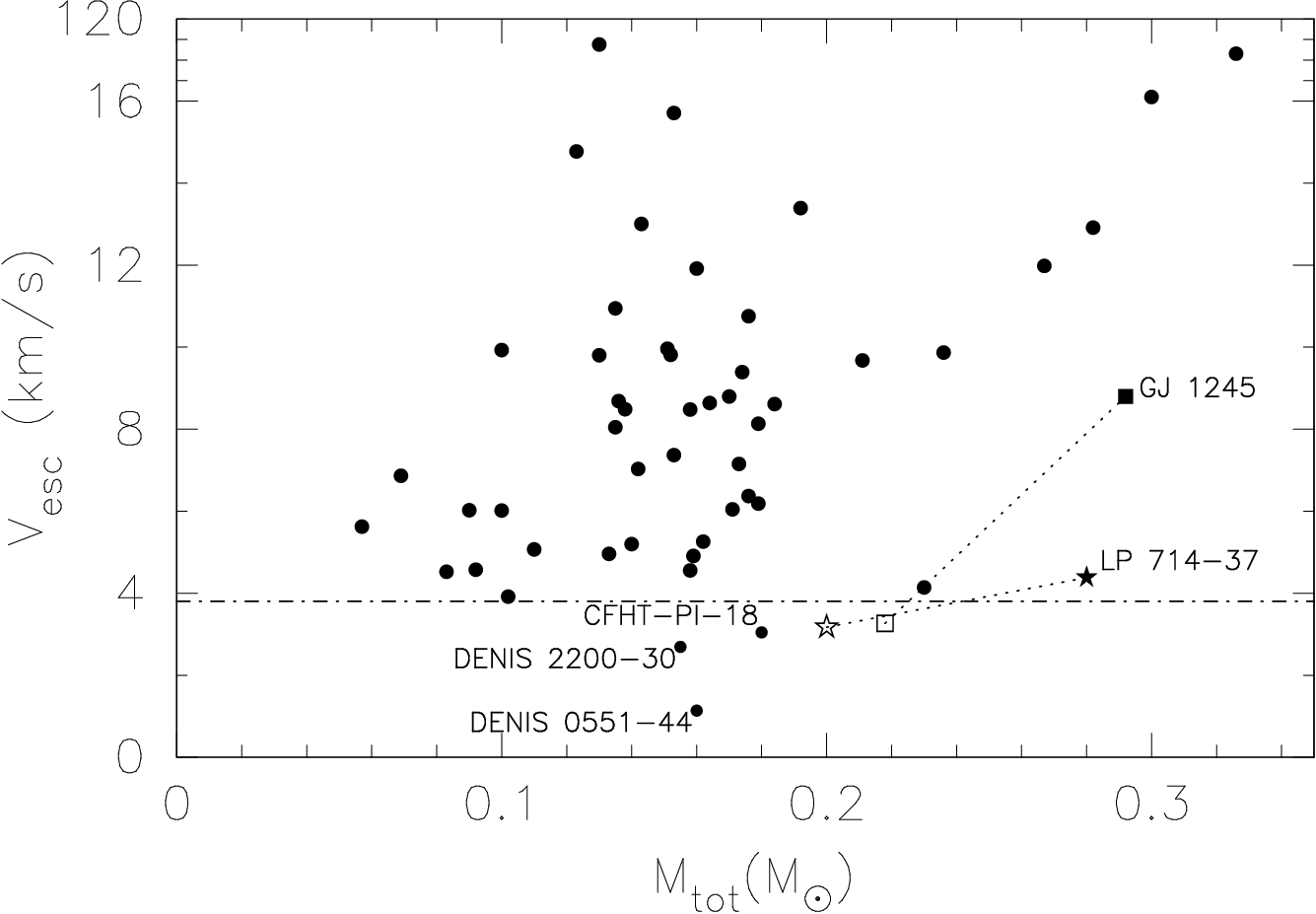}}
\caption{\normalsize 
Escape velocity vs. total mass of the system for VLM systems in the field.
The filled circles represent VLM binaries, from \citet{siegler} and references therein;
\citet{reid97a, beuzit04, martin00, billeres, burgasser06}. 
The squares show the VLM triple system GJ~1245;
the empty square represents the escape velocity calculated by assuming
the system would have only two components A and B meanwhile the filled square 
shows the escape velocity of the triple system.
The LP 714-37 system is noted as different symbols (empty and filled star).
Dash-dotted line: $V_{\rm esc} = 3.8$~km/s \citep{close03}. 
\label{Vesc}}
\end{figure}

\citet{close03} found that the minimum escape velocity in their sample of
34 VLM  binaries is 3.8~km~s$^{-1}$ (Fig.~\ref{Vesc}). The escape
velocity of LP~714-37 A and B at a 33~AU semi-major-axis would be
$V_{\rm esc} \sim 3.3$~km~s$^{-1}$, significantly under that limit.
Accounting for the additional C component however increases the escape
velocity of the system to 4.4~km~s$^{-1}$. This is significantly above the 
3.8~km~s$^{-1}$ \citep{close03} empirical lower limit, and Figure~\ref{Vesc} 
actually has a close analog of LP~714-37: GJ~1245ABC.
LP~714-37 therefore demonstrates that some wide apparent VLM binaries 
are actually higher order multiple systems with 
$V_{\rm esc}~\geq$~$\sim3.8$~km~s$^{-1}$ and no longer contradict the
ejection scenario for brown dwarf formation. The wider VLM binaries
identified by \citet{martin00}, \citet{billeres}, and \citet{burgasser06}
however cannot be brought over $V_{\rm esc}$~$\sim3.8$~km~s$^{-1}$ 
for any realistic number of components. This suggests that other 
formation channels must exist, but a high resolution search for additional 
components among wide VLM binaries such as CFHT-Pl-18 \citep{martin00}, 
DENIS 0551-44 \citep{billeres}, and DENIS 2200$-$3038 \citep{burgasser06}
is clearly important.

\acknowledgments
P.B.N. and E.L.M. acknowledge financial support from NSF grant AST 0440520.
P.B.N. thanks A.J. Burgasser for his comments on the
manuscript. We thank the referee for a fast and useful report. 
This research has made use of the ALADIN, SIMBAD and VIZIER databases, 
operated at CDS, Strasbourg, France.

\clearpage

\clearpage

\clearpage

\clearpage

\clearpage

\begin{table*}
\caption{Relative astrometry and photometry of LP~714-37B and C
relative to LP~714-37A.
  \label{tab_oa}
  }
 $$
\begin{tabular}{lllll} 
 \hline
 \hline
Component & $\rho$  & $\theta$ & $\Delta$(K$^{\prime}$)  \\ 
     &  (arcsec)     & (deg)   &                 \\ 
\hline 
LP~714-37B & 1.87  & 124.9$\pm$0.5 & 0.94$\pm$0.05  \\ 
LP~714-37C & 1.96  & 135.4$\pm$0.5 & 1.24$\pm$0.05  \\ 
 \hline
\end{tabular}
 $$
\end{table*}

\clearpage
\begin{table*}
   \caption{Spectral types of the three components of LP~714-37}
    \label{parameters}
  $$
   \begin{tabular}{llll}
   \hline 
   \hline
   \noalign{\smallskip}
Components    &  SpT        & $M_{\rm K}$ & Mass          \\
              &             &             & ($M_{\odot}$) \\  
    \noalign{\smallskip}
    \hline 
LP 714-37A          &  M5.5$\pm$0.5$^{1}$ &  ~\,9.11$\pm$0.25$^{1}$ &   0.11$\pm$0.01$^{1}$  \\
~~~~~~~~~~~~~\,B    &  M8.0$\pm$0.5       &  10.05$\pm$0.30         &   0.09$\pm$0.01       \\
~~~~~~~~~~~~~\,C    &  M8.5$\pm$0.5       &  10.35$\pm$0.30         &   0.08$\pm$0.01       \\
    \noalign{\smallskip}
    \hline 
   \end{tabular}
  $$
  \begin{list}{}{}
  \item[$^{1}$]: from \citet{phan-bao05} 

{\it Column 1}: Component name.
{\it Column 2}: Spectral type, estimated from the $M_{\rm K}$ versus spectral type relation.
{\it Column 3}: K-band absolute magnitude.
{\it Column 4}: Mass determination for 1-5~Gyr from the models of \citet{baraffe98}.
  \end{list}
\end{table*}
\end{document}